\documentclass[twocolumn,preprintnumbers,showpacs,showkeys,amsmath,amssymb]{revtex4}
\usepackage{graphicx}
\usepackage{dcolumn}
\usepackage{bm}
\begin{document}
\title{Exact Solution of a Reaction-Diffusion Model with Particle Number Conservation}
\author{F. H. Jafarpour $^{1,2}$}
\email{farhad@sadaf.basu.ac.ir}
\author{S. R. Masharian$^1$}
\affiliation{$^1$ Physics Department, Bu-Ali Sina University, Hamadan, Iran}
\affiliation{$^2$ Institute for Studies in Theoretical Physics and Mathematics (IPM), P.O. Box 19395-5531, Tehran, Iran}
\date{\today}
\begin{abstract}
We analytically investigate a 1d branching-coalescing model with
reflecting boundaries in a canonical ensemble where the total number
of particles on the chain is conserved. Exact analytical calculations
show that the model has two different phases which are separated by a
second-order phase transition. The thermodynamic behavior of the canonical
partition function of the model has been calculated exactly in each phase.
Density profiles of particles have also been obtained explicitly. It is shown
that the exponential part of the density profiles decay on three different
length scales which depend on total density of particles.
\end{abstract}
\maketitle
\section{Introduction}
Recently much attention has been payed to the study of shocks in one-dimensional
reaction-diffusion models \cite{fer,bs,kjs,dls}. The shocks are defined as discontinuities
in the space dependence of density of particles in the system and behave as collective
excitations in system. They can be characterized by their
position which performs a random walk. The best known example
in which the shock can appear is asymmetric simple exclusion process
(ASEP) with open boundaries \cite{dehp}. The mathematical relevance of the ASEP is that
it is a discrete version of the Burgers equation in an appropriate scaling limit.
The ASEP contains one class of particles (first class particles) which can be injected
and extracted from the boundaries of a one-dimensional chain while hopping in the bulk
with asymmetric rates. The ASEP has several applications to the realistic systems. For
instance, it can be considered as a simple model for traffic flow \cite{css}. \\
There are different ways to provoke a shock in one-dimensional reaction-diffusion models.
One can consider the ASEP on a closed chain in the presence of a second class particle.
Compared to the first class particles, the second class particles move very slowly. In
\cite{mal,lpk,farhad0,sasamoto,farhad1,farhad2} the shape of the shock is calculated as seen from a second
class particle. Another method is to introduce a slow link in the system \cite{sch2}.
The first class particles cross this link with a smaller crossing rate than that of the other links in
the system. In this case the width of the shock as a function of the length of the
system $L$ scales as $L^{1/3}$ or $L^{1/2}$ depending on whether the density of particles
is equal to $\frac{1}{2}$ or not \cite{jl}. Shocks have also been observed in the ASEP with creation
and annihilation of particles in the bulk of the system \cite{pff,ejs}. \\
In a recent paper we have numerically studied the shocks in a spatially asymmetric
one-dimensional branching-coalescing model with reflecting
boundaries in a canonical ensemble \cite{farhad3}. In this model the particles
diffuse, coagulate and decoagulate on a lattice of length $L$;
however, the total number of the particles is kept fixed. It is
predicted that the model has two different phases and in one phase
the density profile of the particles has a shock structure. We
have confirmed our numerical results by using the Yang-Lee
theory of phase transitions \cite{yanglee} which has recently been shown to be
applicable to the study of critical behaviors of
out-of-equilibrium systems \cite{arndt,be}. In the present work we will show that
by working in the canonical ensemble, the model is exactly solvable in the sense
that the thermodynamic limit of physical quantities can be calculated exactly.
The canonical partition function of the model defined as the sum over stationary
state weights can also be calculated exactly. By applying the Yang-Lee theory we
can calculate the line of the partition function zeros; and therefore, spot the
transition point. The order of the transition can also be identified by investigating
the density of these zeros near the critical point. We will also obtain the exact
expressions for the density profile of the particles on the chain in the thermodynamic
limit. This paper is organized as follows: In Section II we
will define the model and introduce the mathematical preliminaries.
In Section III we will calculate the canonical partition function of
our model using a matrix product formalism and find its behavior in the
thermodynamic limit. We will also find the line of canonical
partition function zeros to confirm our numerical results
in \cite{farhad3}. In Section IV we will calculate the density profile of
the particles on the chain in each phase. In the last section we will discuss
the results and compare them with the case where the total number of particles
is not conserved.
\section{The Model: Mathematical Preliminaries}
In this section we will briefly review the definition of the model
and also define its grand canonical partition function.
We will then calculate the canonical partition function of the model
explicitly. The model consists of one class of particles which diffuse
on a one-dimensional chain of length $L$. Whenever two of these particles
meet, they can coagulate to a single particle. In the same way, a single
particle can decoagulate into two particles. There is no particle input or
output at the boundaries. The reaction rules between two consecutive sites
$i$ and $i+1$ on the chain are explicitly as
follows:
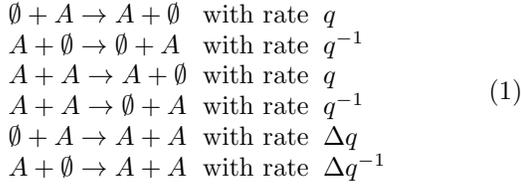
\begin{equation}
\label{Rules}
\begin{array}{ll}
\emptyset+A \rightarrow A+\emptyset & \mbox{with rate} \; \; q \\
A+\emptyset \rightarrow \emptyset+A & \mbox{with rate} \; \; q^{-1} \\
A+A \rightarrow A+\emptyset & \mbox{with rate} \; \; q \\
A+A \rightarrow \emptyset+A & \mbox{with rate} \; \; q^{-1} \\
\emptyset+A \rightarrow A+A & \mbox{with rate} \; \; \Delta q \\
A+\emptyset \rightarrow A+A & \mbox{with rate} \; \; \Delta q^{-1}\\
\end{array}
\end{equation}
in which $A$ and $\emptyset$ stand for the presence of a particle
and a hole respectively. As can be seen, the parameter $q$
determines the asymmetry of the model. For $q>1$ ($q<1$) the particles
have a tendency to move in the leftward (rightward) direction. For any $q$
the model is also invariant under the following transformations:
\begin{equation}
\label{SY} q\longrightarrow q^{-1} \;,\; i\longrightarrow L-i+1
\end{equation}
in which $i$ is a given site on the chain. One should also note
that the rules (\ref{Rules}) do not conserve the number of
particles and therefore the model should be studied in a grand
canonical ensemble. The model without particle number conservation
has already been studied both using Empty Interval Method (EIM)
and Matrix Product Formalism (MPF) \cite{hkp1,hkp2}. It turns out
that the model has two different phases in this case: Two
exponential phases which are called the hight-density and the
low-density phases. On the coexistence line of these phases the
density of the particles on the chain has a linear profile. It is
known that the phase diagram of the ASEP contains a first-order
phase transition line where the injection and the extraction rates
are equal and less than $\frac{1}{2}$. Along this line the density
profile of particles is linear which is a consequence of
superposition of states where a shock between a low-density region
and a hight-density region is present at an arbitrary position
\cite{dehp,andjel}. As we will see the linear profile in present
model can also be interpreted as a sign for the existence of shocks. \\
In order to study the shocks we restrict the total number of
particles on the chain to be $M$ so that their total density is always
equal to $\rho=\frac{M}{L}$. This means that we are working in
a canonical ensemble. The stationary state probability
distribution function can be calculated using the MPF as follows:
We assign two non-commuting operators $D$ and $E$ to a particle
and a hole respectively. Now the probability for occurring any
configuration $\{ \tau \}=\{ \tau_1,\cdots,\tau_L \}$ in the
steady state with exactly $M$ particles can be obtained from
\begin{equation}
\label{DF}
P(\{ \tau \})=\frac{\delta(M-\sum_{i=1}^{L}\tau_i)}{Z_{L,M}}
\langle W \vert \prod_{i=1}^{L}
(\tau_i D + (1-\tau_i) E ) \vert V \rangle
\end{equation}
in which $\tau_i=1$ if the site $i$ is occupied by a particles
and $\tau_i=0$ if it is empty. The normalization factor $Z_{L,M}$
in (\ref{DF}), which will be called the canonical partition function
of the model hereafter, should be obtained from the fact that
$\sum_{ \{ \tau \} }P(\{ \tau \})=1$. It can be writen as
\begin{equation}
\label{CPF} Z_{L,M}=  \sum_{\{\tau\}}
\delta(M-\sum_{i=1}^{L}\tau_i)\langle W \vert\prod_{i=1}^{L}
(\tau_i D+(1-\tau_i)E)\vert V \rangle.
\end{equation}
The Dirac delta in (\ref{DF}) and (\ref{CPF}) guaranties the total number
of particles to be $M$ in the steady state. In order to have a stationary
probability distribution, the operators $D$ and $E$ besides the vectors
$\vert V \rangle$ and $\langle W \vert$ should satisfy the following
quadratic algebra \cite{hkp2}
\begin{equation}
\begin{array}{l}
\label{BulkAlgebra}
[E,\bar{E}] = 0 \\
\bar{E}D-E\bar{D}=q(1+\Delta)ED-\frac{1}{q}DE-\frac{1}{q}D^{2}\\
\bar{D}E-D\bar{E}=-qED+\frac{1+\Delta}{q}DE-qD^2\\
\bar{D}D-D\bar{D}=-q\Delta ED-\frac{\Delta}{q}DE+(q+\frac{1}{q})D^2\\
\langle W \vert\bar{E}=\langle W \vert \bar {D}=0 \; , \; \bar{E}
\vert V \rangle =\bar{D}\vert V \rangle=0.
\end{array}
\end{equation}
The operators $\bar {D}$ and $\bar{E}$ are auxiliary operators and
do not enter into calculating (\ref{DF}) and (\ref{CPF}). The following
four-dimensional representation has been found for the algebra
(\ref{BulkAlgebra}) \cite{hkp2}
\begin{equation}
\label{RepBulkAlgebra}
\begin{array}{c}
D=\left(\begin{array}{cccc}
0&0&0&0\\
0&\frac{\Delta}{1+\Delta}&\frac{\Delta}{1+\Delta}&0\\
0&0&\Delta&0\\
0&0&0&0
\end{array} \right),
\vert V \rangle=\left(\begin{array}{c} a\\0\\q^2\\q^2-1
\end{array} \right),\\
E=\left(\begin{array}{cccc}
q^{-2}&q^{-2}&0&0\\
0&\frac{1}{1+\Delta}&\frac{1}{1+\Delta}&0\\
0&0&1&q^2\\
0&0&0&q^2
\end{array} \right),
\vert W \rangle=\left(\begin{array}{c} 1-q^2\\1\\0\\b
\end{array} \right)
\end{array}
\end{equation}
in which $a$ and $b$ are arbitrary constants and $\vert W \rangle$
is simply transpose of $\langle W \vert$. The matrix
representations for $\bar{D}$ and $\bar{E}$ are also given in \cite{hkp2}.
Using (\ref{RepBulkAlgebra}) one can calculate the steady state
weight of any given configuration.\\
It turns out that the direct calculation of (\ref{CPF}) is not
always an easy task; therefore, we define the grand canonical
partition function which can easily be calculated:
\begin{eqnarray}
\label{GCPF} Z_{L}(\xi) & = & \sum_{\{\tau \}} \langle W
\vert\prod_{i=1}^{L} (\tau_i \xi D+(1-\tau_i)E)\vert V \rangle
\nonumber \\ & = & \sum_{M=0}^{L}\xi^{M} Z_{L,M}
\end{eqnarray}
in which $\xi$ is the fugacity associated with the particles. The
total density of particles $\rho$ should then be fixed by the
fugacity of them through the following equation
\begin{equation}
\label{DFR}
\rho=\lim_{L\rightarrow\infty}\frac{\xi}{L}\frac{\partial}{\partial
\xi} \ln Z_L(\xi).
\end{equation}
One can expect that each value of the fugacity $\xi$ corresponds
to each value of the total density. In this case, the
density-fugacity relation (\ref{DFR}) is invertible and the
equivalence of the canonical and grand canonical ensemble holds.
After calculating the grand canonical partition function (\ref{GCPF}),
one can invert the series to calculate the canonical
partition function using
\begin{equation}
\label{IR} Z_{L,M}=\frac{1}{2\pi i}\int_{C}d\xi\frac{Z_L(\xi)}{\xi^{M+1}}
\end{equation}
where $C$ is a contour which encircles the origin anti-clockwise.
For our model; however, there appears a situation where the
equivalence of ensembles fails in a special region in the
parameters space. There is the place where the shocks appear in the system.\\
As an important physical quantity one can study the density profile of
particles on the chain in the canonical ensemble; nevertheless, the calculation
of the density profile of the particles is much more easily done in the grand
canonical ensemble. Let us define the unnormalized average particle number
at site $i$ in the grand canonical ensemble as:
\begin{widetext}
\begin{equation}
\label{GCDP1}
\langle \rho_i\rangle_L^{(u)}(\xi) = \sum_{\{\tau \}}
\langle W \vert \prod_{j=1}^{i-1}(\tau_j \xi D +(1-\tau_j)E)\xi D
\prod_{j=i+1}^{L}(\tau_j \xi D +(1-\tau_j)E)\vert V \rangle .
\end{equation}
\end{widetext}
We will then translate the results in the grand canonical ensemble
into those in the canonical ensemble using the following formula
\begin{equation}
\label{IRDP} \langle \rho_i \rangle_{L,M}^{(u)} = \frac{1}{2 \pi i}
\int_{C}d\xi \frac{\langle \rho_i\rangle_{L}^{(u)}(\xi)}{\xi^{M+1}}.
\end{equation}
As in (\ref{IR}) the contour $C$ in (\ref{IRDP}) encircles the origin anti-clockwise.
In (\ref{GCDP1}) and (\ref{IRDP}) the superscript $(u)$ means that it is an unnormalized quantity. The 
normalized average particle number at site $i$ should be obtained from 
$\langle \rho_i \rangle=\langle \rho_i \rangle_{L,M}^{(u)}/Z_{L,M}$.
\section{Canonical and Grand Canonical Partition Functions}
In this section we will calculate the grand canonical partition
function of the model explicitly and then using (\ref{IR}) one can
obtain the canonical partition function by applying the steepest
decent method. The grand canonical partition function of this model
can easily be calculated using (\ref{GCPF}) and is simply given by:
\begin{equation}
Z_L(\xi)=\langle W \vert (\xi D+E)^L \vert V \rangle = \langle W \vert C^L \vert V \rangle
\end{equation}
in which we have defined $C:=\xi D+E$. The matrix representations
for the operators $D$ and $E$ are given by (\ref{RepBulkAlgebra}).
After some algebra we find
\begin{equation}
\label{GCPF4}
Z_L(\xi)=Z_L^{(1)}(\xi)+Z_L^{(2)}(\xi)+Z_L^{(3)}(\xi)+Z_L^{(4)}(\xi)
\end{equation}
in which
\begin{widetext}
\begin{eqnarray}
Z_L^{(1)}(\xi)&=&[\frac{-q^4\Delta\xi^2}{(q^2-(1+\xi \Delta))
(q^2(1+\xi \Delta)-1)}](1+\xi\Delta)^L \\
Z_L^{(2)}(\xi)&=&[\frac{q^4(q^2-1)(1+\xi\Delta)}{(q^2+1)(q^2-(1+\xi\Delta))
(q^2(1+\Delta)-(1+\xi \Delta))}]q^{2L}  \\
Z_L^{(3)}(\xi)&=&[\frac{-q^4(q^2-1)(1+\xi\Delta)}{(q^2+1)((1+\Delta)-q^2(1+\xi\Delta))
(1-q^2(1+\xi\Delta))}]q^{-2L} \\
Z_L^{(4)}(\xi)&=&[\frac{q^4\Delta(\xi-1)^2}{(q^2(1+\Delta)-(1+\xi
\Delta))(q^2(1+\xi\Delta)-(1+\Delta))}](\frac{1+\xi\Delta}{1+\Delta})^L.
\end{eqnarray}
\end{widetext}
Because of the symmetry of the model (\ref{SY}) one will only need
to study either the case $q > 1$ or $q < 1$. We will consider the
case $q > 1$ hereafter, and the results for the case $q < 1$ can
easily be obtained by applying the transformations (\ref{SY}).
Obviously for $q > 1 $ we have $q^2 > q^{-2}$. On the other hand
since $\Delta,\xi > 0$ we always have $(1+\xi \Delta) >
(\frac{1+\xi \Delta}{1+\Delta})$. Now two different cases can be
distinguished: We will either have $1 < q < \sqrt{1+\xi \Delta}$
or $1 < \sqrt{1+\xi \Delta} < q $. For these two cases the
asymptotic behaviors of the grand canonical partition function
(\ref{GCPF4}) can be obtained in the large system size limit $L
\rightarrow \infty$:
\begin{equation}
Z_L(\xi)\simeq
\begin{cases}
Z_L^{(1)}(\xi) \;\; ,& \text{$ 1 < q < \sqrt{1 + \xi \Delta} $}  \\
Z_L^{(2)}(\xi) \;\; ,& \text{$ 1 < \sqrt{1+\xi\Delta} < q $}.
\end{cases}
\end{equation}
For a fixed total density of particles $\rho$ (which means fixed $\xi$)
and $\Delta$, the phase transition occurs at $q_c=\sqrt{1+\xi \Delta}$.
Now one can easily calculate the canonical partition function of
the system in these phases using (\ref{IR}). By using (\ref{DFR})
for the first phase the condition $1 < q < \sqrt{1+\xi \Delta}$
translates to $1 < q < \frac{1}{\sqrt{1-\rho}}$ and the canonical
partition function which is given by:
\begin{equation}
\label{ZIa}
Z_{L,M}^{(I)} \simeq \frac{1}{2\pi
i}\int_{C}d\xi\frac{Z_L^{(1)}(\xi)}{\xi^{M+1}}
\end{equation}
can readily be calculated by applying the steepest decent method. We find
\begin{equation}
\label{ZIb} Z_{L,M}^{(I)}\simeq
\frac{q^{4}\Delta^{M-1}\rho^{\frac{3}{2}-M}
(1-\rho)^{M-L-\frac{1}{2}}}{(1-(1-\rho)q^2)(q^2-(1-\rho))} ,
1 < q < \frac{1}{\sqrt{1-\rho}}.
\end{equation}
For the second phase the condition $1<\sqrt{1+\xi\Delta}<q$
translates to $1 < \frac{1}{\sqrt{1-\rho}} < q$. We have also
\begin{equation}
\label{ZIIa}
Z_{L,M}^{(II)} \simeq \frac{1}{2\pi
i}\int_{C}d\xi\frac{Z_L^{(2)}(\xi)}{\xi^{M+1}}.
\end{equation}
Keeping in mind that the contour of the integral above is a unit
circle and that its integrand has two poles, which one of them
$\xi_1=\frac{q^2-1}{\Delta}$ is smaller than unity and the other
$\xi_2=\xi_1+q^2$ is larger than unity, one can easily
calculate (\ref{ZIIa}) using the steepest decent method. We find
\begin{equation}
\label{ZIIb} Z_{L,M}^{(II)}\simeq
\frac{q^{4+2L}\Delta^M}{(q^2+1)(q^2-1)^M} \;\; , \;\;
1 < \frac{1}{\sqrt{1-\rho}} < q.
\end{equation}
It can be seen that for a fixed density $\rho$ the transition
point $q_c=\frac{1}{\sqrt{1-\rho}}$ does not depend on $\Delta$.
This has already been predicted in \cite{farhad3}.
For the case $q < 1$ the transition point is found to be
$q'_c=\sqrt{1-\rho}$ which agrees again with our predications
in \cite{farhad3}. \\
In \cite{farhad3} we have estimated the roots of the canonical partition
function $Z_{L,M}$ as a function $q$ both for $q > 1$ and $q < 1$.
From there we were able to find the transition points. Let us now
calculate the line of the canonical partition function zeros of
the model in the complex $q$-plane for $q > 1$. Defining the
extensive part of the free energy as
\begin{equation}
\label{FE} g=\lim_{L,M\rightarrow\infty}\frac{1}{L}\ln Z_{L,M}
\end{equation}
one can calculate the line of canonical partition function zeros from
\begin{equation}
\label{LZ} Re \; g^{(I)}=Re \; g^{(II)}
\end{equation}
in which $g^{(I)}$ and $g^{(II)}$ are the free energy functions in
the first and the second phase respectively. Using (\ref{ZIb}),
(\ref{ZIIb}), (\ref{FE}) and (\ref{LZ}) we find in the
thermodynamic limit $(L,M\rightarrow\infty,\rho=\frac{M}{L})$:
\begin{equation}
\label{LOZ}
\frac{u^2+v^2}{[(u^2-v^2-1)^2+(2uv)^2]^{\rho/2}}=\frac{(1-\rho)^{\rho-1}}{\rho^\rho}
\end{equation}
in which we have defined $u:=Re(q)$ and $v:=Im(q)$. It can easily
be verified that (\ref{LOZ}) intersects the positive real $q$-axis
at $u_c=\frac{1}{\sqrt{1-\rho}}$. As can be seen the equation (\ref{LOZ})
is exactly the one that we had obtained in \cite{farhad4} for the same model
with the left boundary open and conservation of total number of particles.
In \cite{farhad4} we had also found that the density of canonical partition function
zeros as a function of $q$ drops to zero near the critical point. This indicates
that a second-order phase transition takes place at the critical point.
We have checked that the density of canonical partition function zeros in the
present model also approaches to zero near the critical points $q_c$ and $q'_c$.\\
For $q<1$ we should only change $q \rightarrow q^{-1}$ which means $u \rightarrow
\frac{u}{u^2+v^2}$ and $v\rightarrow \frac{-v}{u^2+v^2}$ in (\ref{LOZ}). In this
case the line of canonical partition function zeros intersects the
positive real $q$-axis at $u'_c=\sqrt{1-\rho}$. It is not difficult
to check that in the thermodynamic limit the numerical estimates
for the canonical partition function zeros obtained in \cite{farhad3} lay exactly on
(\ref{LOZ}) and its counterpart for $q < 1$.
\section{Density Profile of Particles}
Now we consider the average particle number at each site. As for
the partition functions, it turns out that the calculation of
density profile of the particles in the grand canonical ensemble
is much easier than that in the canonical ensemble; therefore, we
will first calculate (\ref{GCDP1}) and then translate out results
into the canonical ensemble using (\ref{IRDP}). The unnormalized
average particle number at site $i$ in the grand canonical
ensemble (\ref{GCDP1}) can also be written as:
\begin{equation}
\label{GCDP2}
\langle \rho_i\rangle_L^{(u)}(\xi)
=\langle W \vert C^{i-1}\xi D C^{L-i} \vert V
\rangle
\end{equation}
in which $C:=\xi D +E$. Now one can use the matrix representation
(\ref{RepBulkAlgebra}) to calculate (\ref{GCDP2}). After some algebra
we find
\begin{widetext}
\begin{eqnarray}
\label{GCDP3}
\langle \rho_i\rangle_L^{(u)}(\xi)  & = & u_1(\xi)
\; q^{2L-4i+2}+u_2(\xi) \; q^{2-2i}(1+\xi\Delta)^{L-i}+
u_3(\xi) \; q^{2-2i}(\frac{1+\xi \Delta}{1+\Delta})^{L-i}+\nonumber \\
& & u_4(\xi) \; q^{2L-2i}(1+\xi \Delta)^{i-1}+u_5(\xi) \;
q^{2L-2i}(\frac{1+\xi \Delta}{1+\Delta})^{i-1}+u_6(\xi)
\; (1+\xi \Delta)^{L-1} + \nonumber \\
& & u_7(\xi) \; (\frac{1+\xi \Delta}{1+\Delta})^{L-1}
\end{eqnarray}
\end{widetext}
in which we have defined
\begin{widetext}
\begin{eqnarray}
u_1(\xi)&=&\frac{q^4(q^2-1)^2\xi \Delta^2 (\xi(2+\xi\Delta)-1)
(q^2-1-\xi\Delta)^{-1}}{(q^2-\xi \Delta -1)(q^2(1+\Delta)-\xi\Delta-1)
(q^2(1+\xi\Delta)-1)(q^2(1+\xi\Delta)-\Delta-1)},\\
u_2(\xi)&=&\frac{-q^2(q^2-1)\xi^2\Delta}{(q^2-\xi\Delta-1)(q^2(1+\xi\Delta)-1)},\\
u_3(\xi)&=&\frac{q^2(q^2-1)(\xi-1)\xi\Delta}{(q^2(1+\Delta)-\xi\Delta-1)(q^2(1+\xi\Delta)-\Delta-1)},\\
u_4(\xi)&=&\frac{q^4(q^2-1)\xi^2\Delta}{(q^2+\xi\Delta-1)(q^2(1+\xi\Delta)-1)},\\
u_5(\xi)&=&\frac{-q^4(q^2-1)(\xi-1)\xi\Delta}{(q^2(1+\Delta)-\xi\Delta-1)(q^2(1+\xi\Delta)-\Delta-1)},\\
u_6(\xi)&=&\frac{-q^4\xi^3\Delta^2}{(q^2-\xi\Delta-1)(q^2(1+\xi\Delta)-1)},\\
u_7(\xi)&=&\frac{q^4(\xi-1)^2\xi\Delta^2}{(1+\Delta)(q^2(1+\Delta)-\xi\Delta-1)(q^2(1+\xi\Delta)-\Delta-1)}.
\end{eqnarray}
\end{widetext}
The asymptotic behaviors of (\ref{GCDP3}) can now be distinguished
for the two mentioned cases. For the first case where $1 < q <
\sqrt{1+\xi \Delta}$ the leading terms are the second, the fourth
and the sixth terms in (\ref{GCDP3}). Now using (\ref{IRDP}) and
(\ref{ZIb}) one can calculate the average particle number of the
particles at site $i$ in the canonical ensemble by applying the
steepest decent method. In the thermodynamic limit the result is:
\begin{equation}
\langle \rho_i\rangle = \rho+(q^2-1)[e^{-\frac{i}{\xi_1}}-
(1-\rho)e^{-\frac{L-i}{\xi_2}}] \; , \;  1 < q <
\frac{1}{\sqrt{1-\rho}}
\end{equation}
in which the correlation lengths are $\xi_1=\vert\ln
(\frac{1-\rho}{q^2}) \vert ^{-1}$ and $\xi_2=\vert
\ln(q^2(1-\rho)) \vert ^{-1}$. For a plot of this profile see
figure 2 in \cite{farhad3}. In the second case where
$1<\sqrt{1+\xi\Delta}<q$ the leading terms are the first and the
fourth terms in (\ref{GCDP3}). Using numerical calculations we had
predicted in \cite{farhad3} that the density profile of the
particles in this phase is a shock in the bulk of the chain while
it increases exponentially near the left boundary for $q > 1$. The
density of the particles in the high-density region of the shock
is equal to $\rho_{High}=1-q^{-2}$ while in the low-density
region, it is zero $\rho_{Low}=0$. One can easily calculate the
share of the first term in to the density profile of the particles
in the canonical ensemble using (\ref{IRDP}). By applying the
steepest decent method one finds $(1-q^{-2})q^{2-4i}$. In order to
calculate the share of the fourth term in (\ref{GCDP3}) in the
grand canonical ensemble we use the following procedure: When $L$
is large, the average density profile can be described by a
continuous function $\rho(x)$ in terms of the rescaled variable
$x=\frac{i}{L}$ where $0 \leq x \leq 1$. By using (\ref{IRDP}) for
the fourth term in (\ref{GCDP3}) we find that the derivative of
$\rho(x)$ has the following form:
\begin{equation}
\label{DPD1}
\frac{d}{dx}\rho(x) \simeq \rho_0 exp[L \cdot F(x)]
\end{equation}
where
\begin{equation}
F(x)=-x\ln q^2+x\ln \frac{x}{x-\rho}-\rho\ln \frac{\rho}{\Delta(x-\rho)}.
\end{equation}
The constant $\rho_0$ in (\ref{DPD1}) is determined by the fact that
\begin{equation}
\int_0^1 \frac{d}{dx}\rho(x) dx=\rho_{Low}-\rho_{High}=q^{-2}-1.
\end{equation}
It turns out that the function $F(x)$ has a maximum value at
$x_0=\frac{\rho}{1-q^{-2}}$. One can expand $F(x)$ around $x_0$ up
to the second order and approximate (\ref{DPD1}) with a Gaussian
and find:
\begin{widetext}
\begin{equation}
\label{DPD2}
\frac{d}{dx}\rho(x) \simeq -\sqrt{\frac{L}{2\pi \rho q^{-2}}} (1-q^{-2})^2
exp(-L\frac{(1-q^{-2})^2(x-x_0)^2}{2\rho q^{-2}}).
\end{equation}
\end{widetext}
By integrating (\ref{DPD2}) the average particle number at site
$i$ in the canonical ensemble for $1 < \frac{1}{\sqrt{1-\rho}} <
q$ is found to be
\begin{widetext}
\begin{equation}
\label{SP}
\langle \rho_i \rangle  = (1-q^{-2})q^2e^{-\frac{i}{\xi_3}}+ \frac{1-q^{-2}}{2}
erfc(\sqrt{\frac{L}{2\rho q^{-2}}}(1-q^{-2})(\frac{i}{L}-\frac{\rho}{1-q^{-2}})) \;,
\; 1 < \frac{1}{\sqrt{1-\rho}} < q
\end{equation}
\end{widetext}
in which the exponential part drops with the length scale
$\xi_3=\vert \ln q^4 \vert^{-1}$ and $erfc(\cdots)$ is the
complementary error function. As can be seen from (\ref{SP}) the
average particle number at site $i$ far from the left boundary is
an error function interpolating between the low-density and the
high-density regions with width scaling as $\sqrt{L}$. For a plot
of this profile see figure 2 in \cite{farhad3}.
\section{Concluding Remarks}
In this paper we studied a one-dimensional asymmetric
branching-coalescing model with reflecting boundaries in a
canonical ensemble where the total number of particles is a
constant. This model has already been studied in literatures in a
grand canonical ensemble where the total number of particles on
the chain is not fixed and can vary between $0$ and $1$ (see \cite{hkp1,hkp2}
and references therein). \\
Without particle number conservation the parameter $\Delta$, which
is the ratio of branching to coalescing rates, governs the average
density of particles on the chain. In this case the phase diagram
of the model consists of two phases: A high-density phase and a
low-density phase. In the hight-density phase the density profile
of the particles has an exponential behavior with two different
correlation lengths $\vert\ln(\frac{q^2}{1+\Delta})\vert^{-1}$ and
$\vert\ln(q^2(1+\Delta))\vert^{-1}$. In the low-density phase the
density profile of the particles has also an exponential behavior;
however, with the length scales $\vert\ln(q^4)\vert^{-1}$ and
$\vert\ln(\frac{q^2}{1+\Delta})\vert^{-1}$. On the coexistence
line between these two phases the density profile of the particles
has a linear decay in one end of the chain while it has an
exponential decay in the other end of the chain with the length scale
$\vert\ln(q^4)\vert^{-1}$. \\
In the canonical ensemble the total density of particles on the
chain is controlled by the parameter $\rho$ instead of $\Delta$.
With particle number conservation it turns out that for $q > 1$
the model has two different phases: An exponential phase and a
shock phase. In the exponential phase the density profile of the
particles has an exponential behavior with two length scales
$\vert\ln(\frac{q^2}{1-\rho})\vert^{-1}$ and $\vert
\ln(q^2(1-\rho)) \vert^{-1}$. In the shock phase the density
profile of the particles drops exponentially near the left
boundary with the length scale $\vert \ln (q^4) \vert^{-1}$. In
the bulk of the chain the density profile of the particles is an
error function with an interface which extends over a region of
width $\sqrt{L}$.

\end{document}